\documentclass[aps,prb,twocolumn,groupedaddress]{revtex4}
\usepackage{graphicx}

\begin{document}


\title{Weak antilocalization in high mobility Ga$_x$In$_{1-x}$As/InP
two-dimensional electron gases with strong spin-orbit coupling}

\author{Vitaliy A. Guzenko}
\email[]{v.guzenko@fz-juelich.de} \affiliation{Institute of Bio- and
Nanosystems (IBN-1) and VISel -- Virtual Institute of Spin
Electronics, Research Centre J\"{u}lich, 52425 J\"{u}lich, Germany}

\author{Thomas Sch\"{a}pers}
\affiliation{Institute of Bio- and Nanosystems (IBN-1) and VISel --
Virtual Institute of Spin Electronics, Research Centre J\"{u}lich,
52425 J\"{u}lich, Germany}

\author{Hilde Hardtdegen}
\affiliation{Institute of Bio- and Nanosystems (IBN-1) and VISel --
Virtual Institute of Spin Electronics, Research Centre J\"{u}lich,
52425 J\"{u}lich, Germany}

\date{\today}

\begin{abstract}

We have studied the spin-orbit interaction in a high mobility
two-dimensional electron gas in a GaInAs/InP heterostructure as a
function of an applied gate voltage as well as a function of
temperature. Highly sensitive magnetotransport measurements of
weak antilocalization as well as measurements of Shubnikov--de
Haas oscillations were performed in a wide range of electron sheet
concentrations. In our samples the electron transport takes place
in the strong spin precession regime in the whole range of applied
gate voltages, which is characterized by the spin precession
length being shorter than the elastic mean free path. The
magnitude of the Rashba spin-orbit coupling parameter was
determined by fitting the experimental curves by a simulated
quantum conductance correction according to a model proposed
recently by Golub [Phys. Rev. B \textbf{71}, 235310 (2005)]. A
comparison of the Rashba coupling parameter extracted using this
model with the values estimated from the analysis of the beating
pattern in the Shubnikov--de Haas oscillations showed a good
agreement.
\end{abstract}


\maketitle

\section{Introduction}
Two-dimensional electron gases (2DEGs) with an InAs or high
In-content Ga$_x$In$_{1-x}$As channel layer are very promising
candidates for spintronic applications, because they show a strong
Rashba spin-orbit interaction along with a high electron
mobility.\cite{Nitta97,Engels97,Schaepers98} These properties are
essential for the realization of various spintronic devices, e.g.,
spin field effect transistors, \cite{Datta90,Schliemann03}
spin-filters, \cite{Bulgakov99,Koga02} or
spin-splitters.\cite{Ohe05} The strength of the Rashba coupling can
be estimated from measurements of Shubnikov--de Haas oscillations by
analyzing the characteristic beating pattern.
\cite{Nitta97,Engels97,Schaepers98} However, the latter can also be
evoked by inhomogeneities of the sheet carrier concentration
\cite{Brosig99,Thillosen06} or due to a slightly occupied second
lowest subband;\cite{Leadley92} thus, the observance of a beating
pattern is not an unambiguous indication of the presence of
spin-orbit coupling. On the other hand, the weak localization effect
is very sensitive not only to an applied magnetic field but also to
spin-orbit coupling. The latter results in a non-monotonous
dependence of the quantum correction to the conductivity on magnetic
field. In case of strong spin-orbit coupling the quantum correction
to the conductivity can even change its sign. The observation of
such an effect, also referred to as weak antilocalization
(WAL),\cite{Hikami80,Bergmann82} is an unambiguous indication of the
presence of spin-orbit interaction. Thus, weak antilocalization
measurements open the way for the experimental determination of the
contributions to the spin splitting: the linear and cubic
Dresselhaus terms,\cite{Dresselhaus55} associated with the lack of
crystal inversion symmetry, and the Rashba term,\cite{Bychkov84}
resulting from the structural inversion asymmetry. The latter can be
controlled by applying an external electric field and is thus in
particular interesting for spintronic
devices.\cite{Nitta97,Engels97}

Until recently, only theoretical models describing the weak spin
precession regime, also called "diffusion" regime, were
available,\cite{ILP94,Pikus95,Knap96} where the elastic mean free
path $l_{tr}$ is much shorter than the spin precession length
$l_{so}$.\cite{Koga_PRL02,Schierholz02}
However, $l_{tr}\gg l_{so}$ is often found in high mobility 2DEGs
comprising a strong spin-orbit interaction. Under such conditions
an experimental observation of the weak antilocalization requires
very sensitive magnetotransport measurements, since the width of
the WAL peak is often less than 1~mT. From theoretical point of
view, extensions of the "diffusion" models to the case of strong
spin precession were only limited to an even narrower range of
magnetic fields and thus do not describe the whole WAL
curve.\cite{Miller03} Very recently, this problem was solved by
the model developed by Golub\cite{Golub05} and later extended by
Glazov and Golub,\cite{Glazov06} which is valid for both the weak
and strong spin precession regimes in 2DEGs.

Taking this into account, we have utilized this model to extract
the spin-orbit coupling in high-mobility GaInAs/InP samples. By
fitting the WAL curves at different gate voltages we obtained the
dependence of the Rashba spin-orbit coupling parameter on the
sheet carrier concentration. The results of these measurements
were compared with the values extracted from the analysis of the
beating pattern of the Shubnikov--de Haas oscillations. We found,
that despite of the fact that the fitting procedure is
time-consuming due to the numerical complexity, the weak
antilocalization measurements might be more advantageous for the
determination of the Rashba coupling parameter, since they are
also applicable if no beating pattern of the Shubnikov--de Haas
oscillations can be observed, e.g. in case of low mobility.

\section{Experimental}
The Ga$_{0.47}$In$_{0.53}$As/Ga$_{0.23}$In$_{0.77}$As/InP
heterostructure used in our investigation was grown by metal
organic vapor phase epitaxy. A 2DEG was formed within the strained
Ga$_{0.23}$In$_{0.77}$As channel layer. A sketch of the layer
sequence is given in Fig.~\ref{alpha_vs_n2D} (inset). Conventional
200-$\mu$m-wide Hall bar structures with voltage probes separated
by 160~$\mu$m were defined by optical lithography and reactive ion
etching. Subsequently, the AuGe ohmic contacts were deposited and
annealed by rapid temperature processing. Metallic top gates
separated from the semiconductor surface by an insulating HSQ
layer (hydrogen silsesquioxane) covered the complete Hall bar
structure and allowed us to control the sheet carrier
concentration $n_{2D}$ in a wide range, even down to a complete
depletion of the 2DEG.

The magnetotransport measurements were performed in a
$^3$He-cryostat with a superconducting magnet at temperatures down
to 0.4~K utilizing a lock-in technique. Since the WAL effect is
strongly temperature dependent, special attention was payed to the
electrical power dissipated in the 2DEG. Depending on the changes of
the resistance of the sample at different gate voltages the ac bias
current was varied to avoid a heating of the 2DEG. In order to
perform weak antilocalization measurements in well-controlled
magnetic fields being less than 15~mT, an additionally mounted small
superconducting coil was used.

\begin{figure}[t]
\includegraphics[width=\columnwidth]{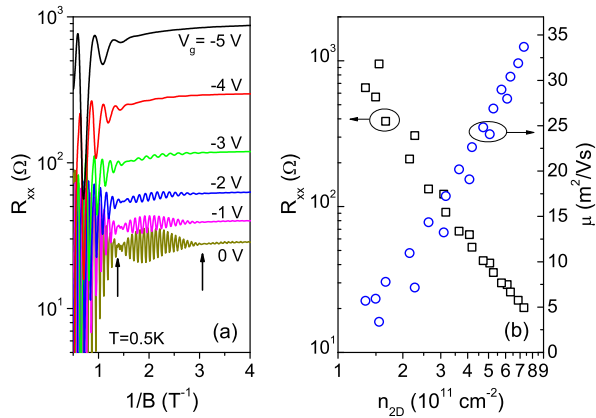}
\caption{\label{Rxx_mu}(Color online) a) Shubnikov--de Haas
oscillations vs. inverse magnetic field measured at different gate
voltages at a temperature of 0.5~K. The nodes of the beating pattern
at zero gate voltage are marked by arrows. The gate voltage was
varied from 0 to -5~V in steps of 1~V. b) Dependence of the
zero-field longitudinal resistance and the electron mobility on the
sheet carrier concentration. The gate voltage range was [0;
-5.9~V].}
\end{figure}
\section{Discussion}
In order to characterize the 2DEG, Shubnikov--de Haas oscillations
were measured in a wide range of gate voltages $V_{g}$ at 0.5~K. As
shown in Fig.~\ref{Rxx_mu}~a), the magnetoresistance vs. inverse
magnetic field curves reveal an increase of the oscillation period
as well as a change of the beating pattern if the gate voltage is
decreased from 0 to $-5$~V. By performing a fast Fourier transform
analysis of the Shubnikov--de Haas oscillations we determined the
corresponding electron sheet densities. No indication of a second
subband occupation was found. As can be seen in Fig.~\ref{Rxx_mu}
b), in the gate voltage range from 0 to -5.9~V we were able to
change the carrier concentration from $7.3\times 10^{11}$~cm$^{-2}$
to $1.3\times 10^{11}$~cm$^{-2}$. On a long-time scale, i.e. within
days, a slight variation of the sheet carrier concentration $n_{2D}$
determined at the same applied gate voltage was observed, possibly
due to the presence of the electron states with a long relaxation
time at the interface between the semiconductor and the gate
insulator. However, the longitudinal resistance $R_{xx}$ as well as
the electron mobility $\mu$ remained unique functions of the carrier
concentration $n_{2D}$ [c.f. Fig.~\ref{Rxx_mu}b)]. As a consequence,
we took $n_{2D}$ rather than $V_{g}$ as a reference for the
following analysis of the weak antilocalization effect.

\begin{figure}[t]
\includegraphics[width=\columnwidth]{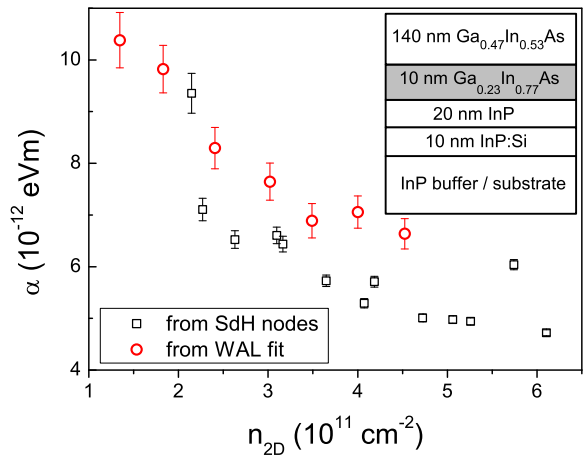}
\caption{\label{alpha_vs_n2D}(Color online) Dependence of the
Rashba coupling parameter $\alpha$ on the sheet carrier
concentration $n_{2D}$ in our GaInAs/InP 2DEG. A good agreement
between the values extracted from the weak antilocalization
measurements (circles) and obtained from analysis of the beating
pattern of the Shubnikov--de Haas oscillations (squares) is
observed. (Inset) A schematic view of the sample cross-section.
2DEG is located within the high In-content GaInAs channel layer.}
\end{figure}

For our heterostructure we assumed that the spin-orbit coupling is
dominated by the Rashba effect.\cite{Schaepers98} Indeed, by
performing self-consistent band structure calculations and applying
the theory presented, e.g. in Ref.~[\onlinecite{Winkler-Book}], the
strength of the Dresselhaus and Rashba spin-orbit coupling at
several gate voltages was estimated. At zero gate voltage, the ratio
between the coupling parameter of the linear Dresselhaus term
($\beta_{calc}=1.1\cdot 10^{-12}$~eVm) and the Rashba term
($\alpha_{calc}=4.6\cdot 10^{-12}$~eVm) is 0.24. With increasing
$V_{g}$ this ratio becomes even less. The cubic Dresselhaus term,
which is dependent strongly on $k_{F}$, at zero gate voltage is
comparable with the linear one and decreases rapidly with reduced
$n_{2D}$. Thus, in our sample the spin-orbit coupling is dominated
by the Rashba effect. By varying the built-in electrical field and,
consequently, the bending of the conductance and valence band
profile the magnitude of the Rashba coupling parameter can be
controlled.

\begin{figure}[tb]
\includegraphics[width=\columnwidth]{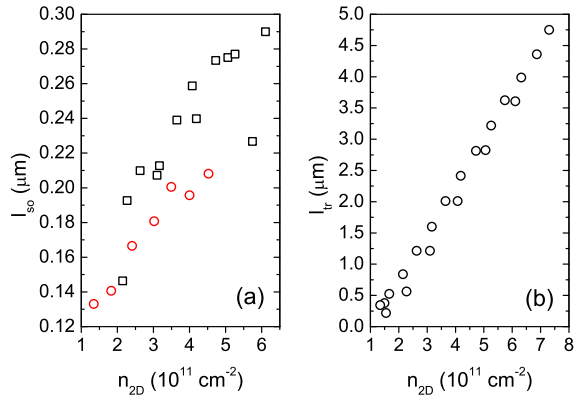}
\caption{\label{lso}(Color online) a) Dependence of the spin
precession length $l_{so}$ on the electron concentration $n_{2D}$;
squares: extracted from the Shubnikov--de Haas measurements;
circles: from weak antilocalization analysis. b) The corresponding
dependence of the mean free path $l_{tr}$.}
\end{figure}

In 2DEGs, the Rashba spin-orbit coupling parameter $\alpha$ can be
determined from the position of the nodes of the beating pattern
in the Shubnikov--de Haas oscillations:\cite{Das89,Schaepers98}
\begin{equation}
\alpha=\frac{\hbar e}{2m^\ast k_{F}}\left(
\frac{1}{B_i}-\frac{1}{B_{i+1}}\right)^{-1},
\end{equation}
where $k_{F}$ is the Fermi wave vector, $m^*$ is the effective
electron mass, and $B_i$ is the magnetic field where the $i$th
node is observed. The values of the spin-orbit coupling parameter
determined by this method are presented in
Fig.~\ref{alpha_vs_n2D}. As can be seen, $\alpha$ increases with
decreasing $n_{2D}$ because of the larger asymmetry of the quantum
well profile at more negative gate voltages. As the sheet carrier
concentration becomes smaller than 2$\times10^{11}$~cm$^{-2}$, the
second node (see Fig.~\ref{Rxx_mu}) cannot be resolved anymore
and, consequently, one faces the limitations of the beating
pattern analysis technique for the extraction of the Rashba
parameter. With the known values of $\alpha$ the spin precession
length can be calculated: $l_{so}=\hbar^2/\sqrt{2}m^\ast \alpha$.
The values of $l_{so}$ at different electron concentrations are
shown in Fig.~\ref{lso}~a). For comparison, the corresponding
values of the elastic mean free path $l_{tr}$ are plotted in
Fig.~\ref{lso}~b).

\begin{figure}[tb]
\includegraphics[width=\columnwidth]{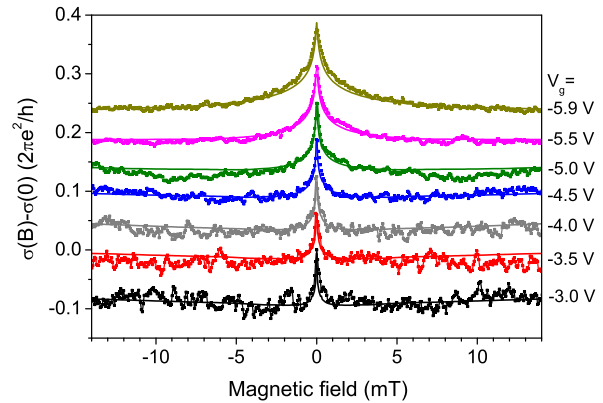}
\caption{\label{WAL-gated}(Color online) Quantum conductivity
correction curves measured at different gate voltages (dots). The
curves are vertically shifted for clarity. Solid lines represent the
fit by a model in Refs.~[\onlinecite{Golub05,Glazov06}]}
\end{figure}

We now turn to the measurements of the weak antilocalization
effect. As can be seen in Fig.~\ref{WAL-gated}, a clear weak
antilocalization peak was resolved at $B=0$ for gate voltages
$V_{g}$ ranging from -3.0 to -5.9~V. Here, the magnitude of the
quantum conductance correction was obtained by subtracting the
conductance at zero magnetic field from the experimental
determined magnetoconductance. \endnote{This is acceptable because
the change of the magnetoconductance due to the Lorentz force,
which is a classical effect, is negligible in the range of
magnetic field applied here.} Increasing the gate voltage resulted
in a strong increase of the sample resistance and simultaneously
in a broadening of the WAL peak. Under such conditions, the
experimental observation of the WAL effect becomes easier,
compared to resolving the beating pattern of the Shubnikov--de
Haas oscillations, in particular at the highest gate voltages. At
$V_{g}\leq-2.5$~V the WAL effect could not be resolved
unambiguously.

In order to extract the Rashba coupling parameter from the weak
antilocalization measurements the experimental curves were fitted by
numerically calculated ones. The choice of the proper model is
governed by the transport regime in the sample. In our case, a
comparison of the elastic mean free path $l_{tr}$ and the spin
relaxation length $l_{so}$ extracted from the findings of the
Shubnikov--de Haas oscillations revealed that $l_{tr}$ is larger
than $l_{so}$. This corresponds to the regime of strong spin
precession, so that the model in
Refs.~[\onlinecite{Golub05,Glazov06}] has to be applied. It has been
shown that the calculated weak antilocalization curves do not differ
from each other significantly if the ratio between the coupling
parameters $\beta$ and $\alpha$ is less than approximately
0.6.\cite{Glazov06} Since in our case the ratio is less than 0.24,
we can readily neglect the Dresselhaus terms in the further
analysis. At a given electron concentration, the number of free
fitting parameters in the simulations could be reduced from four to
two, since the elastic mean free path $l_{tr}$ and the elastic
scattering time $\tau_{tr}$ were determined directly from the
Shubnikov--de Haas measurements and were kept constant during the
fitting procedure. The initial value of one of the free parameters,
the phase coherence length $l_{\varphi}$, was estimated according to
Refs.~[\onlinecite{Altshuler82,Giuliani82,Fukuyama83,Choi87}] and
adjusted during the fitting procedure. As can be seen in
Fig.~\ref{WAL-gated}, a good fit to the experimental curves has been
achieved for all gate voltages. The corresponding values of $\alpha$
are shown in Fig.~\ref{alpha_vs_n2D}.

Knowing the fitting parameters in particular the dependence of
$\alpha$ on $n_{2D}$, the corresponding dependence of the spin
precession length $l_{so}$ was determined [see Fig.~\ref{lso}~a)].
High electron concentrations, i.e. small negative gate voltages,
result in $l_{so}$ which is an order of magnitude shorter than
$l_{tr}$. Applying higher negative gate voltages, i.e. lowering
$n_{2D}$, leads to the rapid shortening of the $l_{tr}$ and, due
to the enhancement of the spin-orbit interaction, to the
shortening of $l_{so}$. Despite the decrease of $l_{tr}$ with
$n_{2D}$, $l_{tr}$ is always larger than the corresponding value
of $l_{so}$. Thus the strong spin precession regime of the
electron transport is preserved in the whole range of electron
concentrations studied here.
\begin{figure}[tb]
\includegraphics[width=\columnwidth]{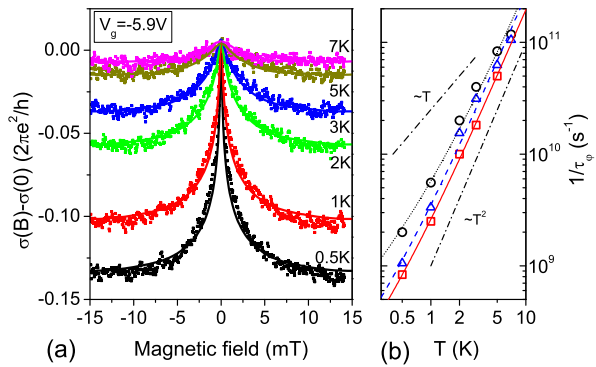}
\caption{\label{WAL-Temp}(Color online) a) Quantum conductivity
correction curves measured at different temperatures with a constant
gate voltage (dots). Solid lines represent the fit by a model in
Refs.~[\onlinecite{Golub05,Glazov06}]. b) Temperature dependence of
the phase breaking rate $1/\tau_{\varphi}$ extracted from the fits
of the WAL curves measured at gate voltage of -5.9~V (circles),
-5.5~V (triangles) and -4.5~V (squares). The lines represent the
best fit by a function taking into account a $T$ and $T^2$
dependence of $1/\tau_\phi$.\cite{Choi87} The dashed-dotted lines
indicate slopes corresponding to $T$ and $T^2$.}
\end{figure}

A comparison of the values of $\alpha$ obtained by the analysis of
the beating pattern with the ones determined from the WAL
measurements reveals a good agreement, although the latter are
slightly higher. The origin of this discrepancy is not clear, yet.
Possibly, the inaccuracy in the determination of the minimum
position of the WAL curve caused by the additional contribution of
other magnetotransport effects results in an overestimation of the
Rashba coupling parameter. As long as the spin-orbit interaction
is the only cause of the beatings in the Shubnikov--de Haas
oscillations, these two methods are complementary in a certain
range of carrier densities. However, at high carrier densities the
observation of WAL is more difficult, since the longitudinal
resistance of the sample and consequently the signal-to-noise
ratio drops rapidly, whereas the characteristic beating pattern in
the Shubnikov--de Haas oscillations can be resolved easily. In
contrast, at low electron concentrations the WAL peak can be
easily measured, whereas the position of the nodes cannot be
resolved anymore.

Temperature dependent measurements, presented in
Fig.~\ref{WAL-Temp}~a) for a gate voltage of $-5.9$~V, show a fast
suppression of the WAL peak with increasing temperature. Also
here, a good agreement between the experiment and the model of
Glazov and Golub\cite{Glazov06} can be readily seen. The only
parameter which was varied during the fitting procedure was the
phase coherence time $\tau_{\varphi}$ [see
Fig.~\ref{WAL-Temp}~b)].

At low temperatures the phase breaking rate $1/\tau_{\varphi}$ is
essentially determined by electron-electron scattering, which can
be divided into two contributions. The first one, being connected
to a large-energy-transfer scattering mechanism, leads to a
$T^2$-dependence of $1/\tau_\phi$ and is dominant at $T>\hbar
/k_{B}\tau_{tr}$.\cite{Fukuyama83,Giuliani82,Zheng96} The second
contribution originating from a small-energy-transfer mechanism
depends linearly on $T$ and is most significant at $T<\hbar
/k_{B}\tau_{tr}$.\cite{Altshuler82,Fukuyama83} As can be seen in
Fig.~\ref{WAL-Temp}~b), the experimental points of the phase
breaking rate could be fitted well to a combination of $T$ and
$T^2$ dependencies. A comparison with the slopes corresponding to
$T$ and $T^2$ [cf. Fig.~\ref{WAL-Temp}b)] confirms that at higher
temperatures ($T
>1$~K) large-energy-transfer scattering with $1/\tau_\phi \propto
T^2$ dominates. Whereas, at lower temperatures a deviation from a
slope proportional to $T^2$ towards a linear temperature
dependence is observed, in particular at $V_g=-5.9$~V. In fact, at
increasing negative gate voltages a shift of the crossover
temperature $\hbar /k_{B}\tau_{tr}$ towards larger values is
expected, owing to the decrease of $\tau_{tr}$ with decreasing
electron concentration. In detail, at gate voltages of $-5.9$~V,
$-5.5$~V, and $-4.5$~V the cross-over temperature was determined
to be 5.4~K, 3.9~K, and 2.4~K, respectively. In accordance with
the theoretical prediction
\cite{Altshuler82,Giuliani82,Fukuyama83,Zheng96} an increased
scattering rate $1/\tau_\phi$ is found for larger negative gate
voltages, i.e. lower $n_{2D}$ and lower $\tau_{tr}$. However, a
direct comparison of the values of $1/\tau_\phi$ extracted from
the fit to the Golub model\cite{Golub05} with the theoretically
determined values\cite{Choi87} reveals, that the latter is larger
by a factor of about two. Probably, this discrepancy is connected
to uncertainties in the determination of $\tau_\phi$ at large
negative gate voltages, which is caused by the weak dependency of
the quantum conduction correction on $\tau_\phi$. As we observed
in our calculations, $\tau_\phi$ has a strong effect on the
magnitude of the quantum correction at magnetic fields smaller
then 0.1~mT, i.e. close to the resolution limit of the experiment.
In addition, inaccuracies in the determination of $l_{tr}$ and
$\tau_{tr}$ from the Shubnikov--de Haas oscillations become more
pronounced at large negative gate voltages. In our simulations we
found that these inaccuracies mainly affect the precision in the
determination of $\tau_\phi$.

\section{Conclusions}
In conclusion, we have studied weak antilocalization in a GaInAs/InP
2DEG as a function of electron concentration. Experimental curves
were fitted by a universal model which describes both the weak and
the strong spin precession regime of the electron transport.
Satisfactory fits were achieved in a wide range of gate voltages as
well as at different temperatures. We have shown, that the
dependence of spin-orbit coupling on the gate voltage can be
successfully studied in high mobility 2DEGs by analyzing weak
antilocalization measurements. This is a reliable method which is
complementary to the beating pattern analysis of the Shubnikov--de
Haas oscillations.

\acknowledgments The authors are deeply grateful to L. E. Golub and
M. M. Glazov (A. F. Ioffe Physico-Technical Institute, St.
Petersburg, Russia) for fruitful discussions.


\end{document}